\documentclass[12pt,prd,aps,amssymb,amsmath,tightenlines,showpacs,a4paper,nofootinbib]{revtex4}
\def\bea{\begin{eqnarray}}
\def\eea{\end{eqnarray}}
\def\beax{\begin{eqnarray*}}
\def\eeax{\end{eqnarray*}}
\def\half{\frac{1}{2}}
\def\quarter{\frac{1}{4}}
\def\vf{\varphi}
\usepackage{graphicx}

\begin{document}
\title{$PT$-Symmetric Sinusoidal Optical Lattices at the Symmetry-Breaking Threshold}

\author{Eva-Maria~Graefe$^\dag$\email{e.m.graefe@imperial.ac.uk} and H.~F.~Jones$^\ddag$\email{h.f.jones@imperial.ac.uk}}

\affiliation{
$\phantom{.}^\dag$Mathematics Department, Imperial College, London SW7 2BZ, UK\\
$\phantom{.}^\ddag$Physics Department, Imperial College, London SW7 2BZ, UK\\}
\date{\today}

\begin{abstract}
The $PT$ symmetric potential $V_0[\cos(2\pi x/a)+i\lambda\sin(2\pi x/a)]$ has a completely real spectrum for $\lambda\le 1$, and begins to develop complex eigenvalues for $\lambda>1$. At the symmetry-breaking threshold $\lambda=1$ some of the eigenvectors become degenerate, giving rise to a Jordan-block structure for each degenerate eigenvector. In general this is expected to result in a secular growth in the amplitude of the wave. However, it has been shown in a recent paper by Longhi, by numerical simulation and by the use of perturbation theory, that for a broad initial wave packet this growth is suppressed, and instead a saturation leading to a constant maximum amplitude is observed. We revisit this problem by explicitly constructing the Bloch wave-functions and the associated Jordan functions and using the method of stationary states to find the dependence on the longitudinal distance $z$ for a variety of different initial wave packets. This allows us to show in detail how the saturation of the linear growth arises from the close connection between the contributions of the Jordan functions and those of the neighbouring Bloch waves.

\end{abstract}

\pacs{42.25.Bs, 02.30.Gp, 11.30.Er, 42.82.Et}
\maketitle

\section{Introduction}
The study of quantum mechanical Hamiltonians that are $PT$-symmetric but not Hermitian\cite{BB}-\cite{AMh} has recently found an unexpected application in classical optics\cite{op1}-\cite{op9}, due to the fact that in the paraxial approximation the equation of propagation of an electromagnetic wave in a medium is formally identical to the Schr\"odinger equation, but with different interpretations for the symbols appearing therein. The equation of propagation takes the form
\bea\label{opteq}
i\frac{\partial\psi}{\partial z}=-\left(\frac{\partial^2}{{\partial x}^2}+V(x)\right)\psi,
\eea
where $\psi(x,z)$ represents the envelope function of the amplitude of the electric field, $z$ is a scaled propagation distance, and $V(x)$ is the optical potential, proportional to the variation in the refractive index of the material through which the wave is passing. A complex $V$ corresponds to a complex refractive index, whose imaginary part represents either loss or gain. In principle the loss and gain regions can be carefully configured so that $V$ is $PT$ symmetric, that is $V^*(x)=V(-x)$.

Propagation through such a medium exhibits many new and interesting properties, such as nonreciprocal diffraction \cite{Berry98} and birefringence \cite{op7}. One of the main features of complex optical lattices is the non-conservation of the total power. In the $PT$-symmetric case this can lead to effects such as power oscillations \cite{op7}. It has been argued that one can distinguish three universal dynamics \cite{Kottos_uni} related to broken or unbroken symmetry. While this is in general true, the behaviour can be modified considerably for special initial conditions, as we will discuss in the present paper. Many familiar effects such as Bloch oscillations and dynamical localisation get drastically modified in the presence of imaginary potentials and $PT$-symmetry \cite{Longhi_bloch, Longhi_dl}. The new features of complex optical lattices provide exciting opportunities for engineering applications. As an example, the possibility of realizing unidirectional light propagation has been envisaged \cite{Kulishov}. In the case of high intensities the propagation equation (\ref{opteq}) gets modified due to the Kerr-nonlinearity, leading to an additional term proportional to $|\psi|^2\psi$. It has been shown in \cite{op9} that the influence of the nonlinearity on the non-reciprocal effects can be advantageous for applications such as unidirectional couplers. It is interesting to note that the nonlinear propagation equation also has a counterpart in quantum dynamics, as the mean-field description of Bose-Einstein condensates, where there has also been interest in $PT$ symmetric models \cite{nhbh}. However, for the purposes of this paper, we shall limit ourselves to the linear case.

A model system exemplifying some of the novel features of beam propagation in $PT$-symmetric optical lattices uses the sinusoidal potential
\beax\label{Vsin}
V=V_0\left[\cos(2\pi x/a)+i \lambda \sin(2\pi x/a)\right]\ .
\eeax
This model has been studied numerically and theoretically, e.g. in Refs.~\cite{op3,op6,op7}. The propagation in $z$ of the amplitude $\psi(x,z)$ is governed by the analogue Schr\"odinger equation (\ref{opteq}), which for an eigenstate of $H$, with eigenvalue $\beta$ and $z$-dependence $\psi\propto e^{-i\beta z}$ reduces to the eigenvalue equation
\bea\label{H}
-\psi''-V_0\left[\cos(2\pi x/a) + i \lambda \sin(2\pi x/a)\right]\psi= \beta\psi\ .
\eea
 These eigenvalues are real for $\lambda\le 1$, which corresponds to unbroken $PT$ symmetry, where the eigenfunctions respect the (anti-linear) symmetry of the Hamiltonian. Above $\lambda=1$ pairs of complex conjugate eigenvalues begin to appear, and indeed above $\lambda\approx 1.77687$ all the eigenvalues are complex\cite{Midya}. Clearly one would expect oscillatory behaviour of the amplitude below the threshold at $\lambda=1$ and exponential behaviour above the threshold, but the precise form of the evolution at $\lambda=1$ is less obvious. At first sight one would expect linear growth (see, e.g. Ref.~\cite{Heiss}) because of the appearance of Jordan blocks associated with the degenerate eigenvalues that merge at that value of $\lambda$ , but, as Longhi\cite{op6} has emphasized, this behaviour can be significantly modified depending on the nature of the initial wave packet.

It is this problem that we wish to discuss in the present paper. In Section 2 we explicitly construct the Bloch wave-functions and the associated Jordan functions corresponding to the degenerate eigenvalues and then use the analogue of the method of stationary states to construct the $z$-dependence. We find that the explicit linear dependence arising from the Jordan associated functions is indeed cancelled by the combined contributions from the non-degenerate wave-functions (which individually give an oscillatory behaviour). In Section 3 we analyze this cancellation in detail, showing how the coefficients of the two contributions are closely related, and obtaining an approximate analytic expression for the $z$-derivative of their sum. Our conclusions are given in Section 4.
\section{Bloch and associated Jordan wave-functions}
At the threshold $\lambda=1$, the potential $V$ in Eq.~(\ref{Vsin}) becomes the complex exponential $V=V_0 \exp(2i\pi x/a)$, for which the Schr\"odinger equation reads
\bea\label{H1}
-\psi''-V_0\exp(2i\pi x/a)\psi= \beta\psi .
\eea
This is a form of the Bessel equation, as can be seen by the substitution $y=y_0 \exp(i \pi x/a)$, where $y_0=(a/\pi)\surd V_0$, giving
\bea\label{Bessel}
y^2\frac{d^2\psi}{dy^2}+y\frac{d\psi}{dy}-(y^2+q^2)\psi=0,
\eea
where $q^2=\beta(a/\pi)^2$. Thus the spectrum is that of a free massive particle, shown in the reduced zone scheme in Fig.~1, and for $q\equiv ka/\pi$ not an integer the solutions $\psi_k(x)=I_q(y)$ and $\psi_{-k}(x)=I_{-q}(y)$ are linearly independent, and have exactly the correct periodicity, $\psi_k(x+a)=e^{ika}\psi_k(x)$, to be the Bloch wave-functions. It is important to note, however, that because the original potential is $PT$-symmetric rather than Hermitian, these functions are not orthogonal in the usual sense, but rather with respect to the $PT$ inner product (see Eq.~(\ref{orthogonality})).
\begin{figure}[h]
\resizebox{!}{3in}{\includegraphics{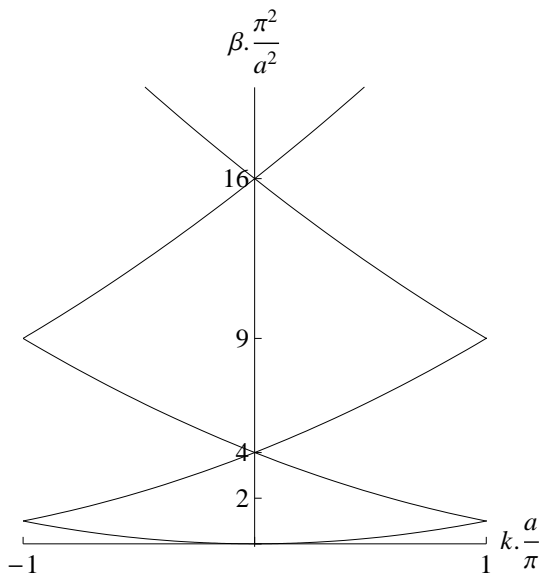}}
\caption{Band structure for $\lambda=1$ in the reduced zone scheme. The Bloch momentum $k$ is plotted in units of $\pi/a$ and the eigenvalue $\beta$ in units of $(a/\pi)^2$.}
\end{figure}
\subsection{Jordan Associated Functions}
However, for $q=n$, a non-zero integer, $I_n(y)$ and $I_{-n}(y)$ are no longer independent, but are in fact equal, signalling the degeneracy of the eigenvectors at those points, and the formation of spectral singularities and Jordan blocks. In that case the Bloch eigenfunctions do not form a complete set, and we must search for other functions, still with the same periodicity, to supplement them. These are the Jordan associated functions (see Appendix and Refs.~\cite{Kato,Keldysh} ), which we denote by $\vf_k(x)\equiv \chi_n(y)$, defined not by the eigenvalue equation itself, but by
\bea\label{Jordan}
\left[y^2\frac{d^2}{dy^2}+y\frac{d}{dy}-(y^2+n^2)\right]\chi_n(y)=I_n(y),
\eea
and the periodicity condition $\chi_n(e^{i\pi}y)=e^{i\pi}\chi_n(y)$, corresponding to the
Bloch periodicity $\vf_k(x+a)=e^{ika}\vf_k(x)$.
A particular solution of this equation, which can be expressed explicitly in terms of generalized hypergeometric functions, is given by
\bea\label{particular}
\chi_n^{PI}(y)=\int_{y_0}^y \frac{dz}{z} I_n(z)\left[K_n(z)I_n(y)-K_n(y)I_n(z)\right],
\eea
as is easily checked by differentiation and use of the Wronskian identity $K_n(y)I_n'(y)-K_n'(y)I_n(y)=1$. However, the corresponding $\vf_k^{PI}(x)$ has a discontinuity in its imaginary part at $x=\pm a$ and does not have the required periodicity. This problem can be rectified by recognizing that we may add to $\chi_n^{PI}$ any multiple of $I_n(y)$, or more importantly $K_n(y)$. The latter displays exactly the same kind of discontinuity\footnote{This arises because $K_n(y)$ has a branch cut along the negative real axis, and when $x$ passes through $\pm a$ one should really go to the next Riemann sheet.} as $\chi_n^{PI}$, and by choosing its coefficient judiciously the discontinuities can be made to cancel. Specifically, we take
\bea\label{fixed}
\chi_n(y)=\chi_n^{PI}(y)-\half K_n(y).
\eea
An alternative derivation of this relation using the definition of $\vf_k(x)$ in terms of $d\phi_k(x)/dk$ will be given in the next section.
With the addition of the last term the resulting $\vf_k(x)$ is not only free of discontinuities, but also obeys the correct Bloch periodicity condition. This is shown in Figure 2, where we plot the real and imaginary parts of $\vf_k^{PI}(x)$ and $\vf_k(x)$ for $q=2$. Note the $PT$ symmetry of $\vf_k(x)$, namely $\vf_k(-x)=\vf_k^*(x)$.
\begin{figure}[h!]
\resizebox{!}{4.5cm}{\includegraphics{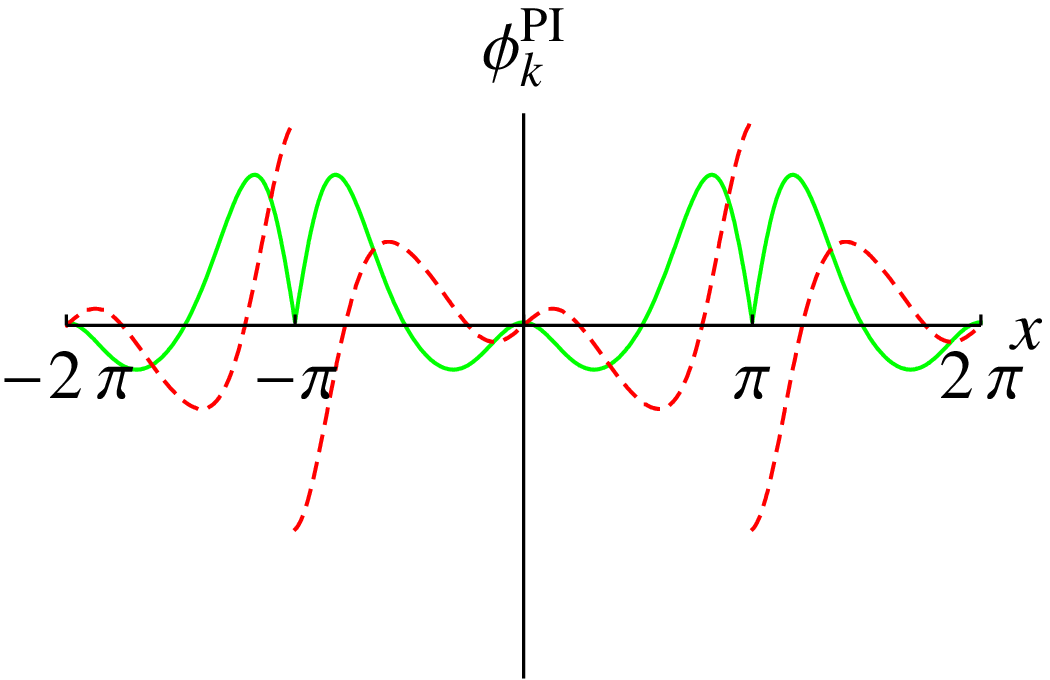}}\hspace{1cm}\resizebox{!}{4.5cm}{\includegraphics{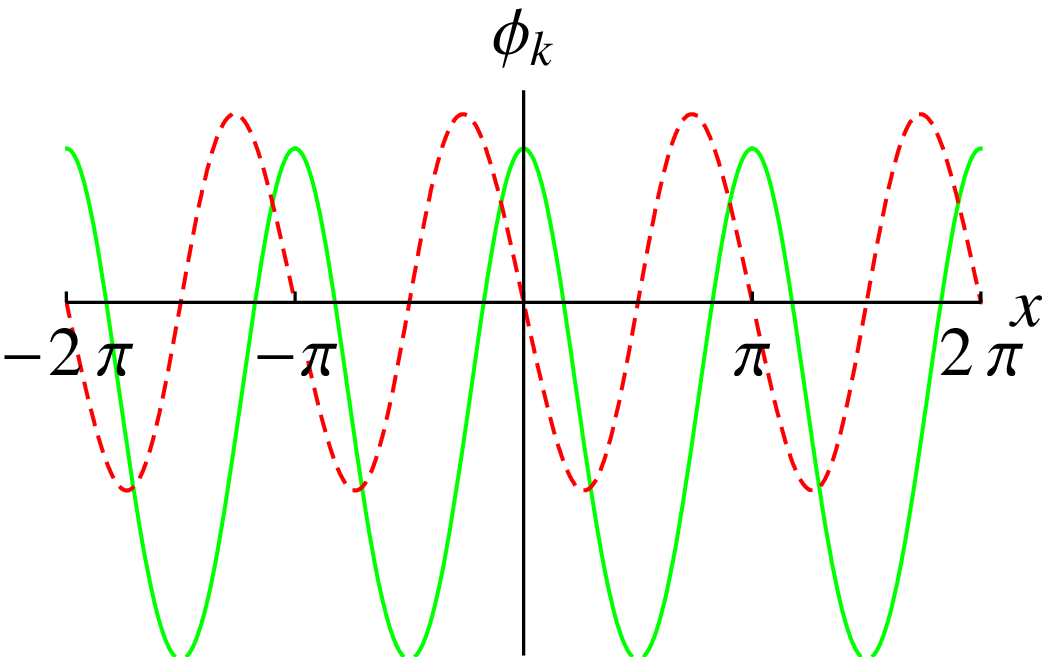}}
\centerline{(a) \hspace{6.5cm} (b)}
\caption{(color online) The real (green, continuous) and imaginary (red, dashed) parts of (a) the particular integral $\vf_k^{PI}(x)$ and (b) the corrected Jordan function $\vf_k(x)$ for $a=\pi$, $k=q=2$ and $V_0=2$.}
\end{figure}

The complete set of functions, orthogonal with respect to the $PT$ metric, now consists of the Bloch eigenfunctions $\psi_k(x)$, supplemented by the Jordan associated functions $\vf_k(x)$ for $k=n\pi/a$ with $n>0$, and a general wave-function $f(x)$ may be expanded as
\bea\label{PTexpansion}
f(x)=c_0 \psi_0(x)+\sum_{k\ne n\pi/a} c_k \psi_k(x)+\sum_{n >\ 0}(\alpha_n I_n(y)+\beta_n \chi_n(y)).
\eea
Here we have discretized the problem by putting the system in a box of length $2Na$, in which case $k\to k_r= r\pi/(N a)$.
The coefficients $c_k$ can be obtained\cite{op3}
by using the $PT$ orthogonality
\bea\label{orthogonality}
\int dx \psi_{-k}(x)\psi_{k'}(x)=\delta_{k k'}\int dx \psi_{-k}(x)\psi_k(x),
\eea
as
\bea\label{PTcoeffs}
c_k=\frac{\int dx \psi_{-k}(x) f(x)}{\int dx \psi_{-k}(x)\psi_k(x)}\ .
\eea
Here the sign of the denominator alternates from band to band, reflecting the indefinite nature of the $PT$ metric. However, this relation breaks down precisely at the Brillouin Zone boundaries, where the single Bloch eigenfunctions are self-orthogonal, $\int dx\ I_n^2(y)=0$, another indication that we need the supplementary Jordan functions.

Thus it is $\beta_n$, rather than $\alpha_n$, which is determined by integrating $f(x)$ with respect to $I_n(y)$:
\bea\label{Jordancoeff}
\beta_n = \frac{\int dx I_n(y) f(x)}{\int dx I_n(y) \chi_n(y)},
\eea
while $\alpha_n$ is subsequently determined by integrating $f(x)$ with respect to $\chi_n(y)$.

As an immediate check of the correctness of the functions $\chi_n$, the identity
\bea\label{one}
1=I_0(y)-2I_2(y)+2I_4(y)-\dots
\eea
implies that
\bea
\int dx \chi_{2n}(y)=2(-1)^n \int dx I_{2n}(y)\chi_{2n}(y),
\eea
for $n>0$, a relation we have verified numerically.

\subsection{Method of Stationary States}
In standard quantum mechanics one method of solving the time-dependent Schr\"odinger equation
\bea
i\frac{d}{dt}\psi(x,t)=H \psi(x,t)
\eea
is to expand the initial wave-function $\psi(x,0)$ in terms of the (complete) set of orthonormal eigenfunctions $\psi_r$ as
\bea
\psi(x,0)=\sum_m a_m \psi_m(x),
\eea
with the coefficients $a_m$ given by the overlap
\bea
a_m=\int dx \psi^*_m(x)\psi(x,0).
\eea
The wave-function at time $t$ is then given by
\bea
\psi(x,t)=\sum_m a_m e^{-iE_m t} \psi_m(x).
\eea
Here we have essentially the same problem, with $z$ taking the role of $t$, but with the crucial difference that we must include the Jordan associated functions in the sum, which now takes the form of Eq.~(\ref{PTexpansion}), with the coefficients determined as in Eqs.~(\ref{PTcoeffs}) and (\ref{Jordancoeff}).  As is well known, and has been emphasized again recently in Refs.~\cite{op6} and\cite{Heiss}, the time dependence then takes a different form. The Jordan function $\vf_r$ satisfies the equation $(H-E_r)\vf_r=\psi_r$, and hence
\bea
e^{-iHt}\vf_r &=& e^{-iE_r t} e^{-i(H-E_r)t} \vf_r\cr
&=& e^{-iE_r t}(\vf_r -i t \psi_r).
\eea
Thus, in addition to the usual phase factors making up the sum for the time-dependent wave-function, one has an explicit factor of $t$ multiplying the degenerate eigenfunctions associated with a Jordan block, giving the complete time dependence of the initial wave-function of Eq.~(\ref{PTexpansion}) as
\bea\label{J+rest}
f(x,t)&=&c_0 \psi_0(x)+\sum_{k\ne n\pi/a} c_k \psi_k(x)e^{-ik^2t}\nonumber\\
&&\hspace{1cm}+\sum_{n >\ 0}((\alpha_n -it\beta_n)I_n(y)+\beta_n \chi_n(y))e^{-i(n^2\pi^2/a^2)t}.
\eea
However, the explicit factor of $t$ only appears when the coefficient of the Jordan function is nonzero. The expansion of Eq.~(\ref{one}) is a case in point.

In the corresponding optical problem it therefore appears that the $z$-dependence should be expected to be oscillatory for $\lambda<1$, exponential for $\lambda>1$ and linear precisely at the symmetry-breaking threshold $\lambda=1$ for initial states that excite a Jordan function. However, Longhi\cite{op6} has shown numerically and in perturbation theory that this linear dependence may not be realized, depending on the nature of the initial wave-function. Since we now have explicit expressions for the Bloch and Jordan functions we are in a position to investigate the origin
of this phenomenon in the context of the method of stationary states.

We will take as our input a Gaussian profile of the form
\bea\label{Gaussian}
\psi(x,0)=f(x)\equiv e^{-(x/w)^2+i k_0 x},
\eea
with offset $k_0$ and width $w$. Because of the periodicity of the Bloch eigenfunctions and Jordan associated functions the range of the integral in Eqs.~(\ref{PTcoeffs}) and (\ref{Jordancoeff}) can be reduced to $0\le x\le a$, provided that $f(x)$ is replaced by
\bea\label{FN}
F_q(x)\equiv \sum_{m=-N}^{N-1} e^{-i\pi m q}f(x+ma),
\eea
in which we recall that  $q=ka/\pi$. That is,
\bea
c_k=\frac{\int_0^a dx \psi_{-k}(x) F_q(x)}{2N\int_0^a dx \psi_{-k}(x)\psi_k(x)},
\eea
and similarly for $\alpha_n$, $\beta_n$. In particular
\bea\label{InIntegral}
\beta_n = \frac{\int_0^a dx I_n(y) F_{n}(x)}{2N\int_0^a dx I_n(y) \chi_n(y)}.
\eea
For $N$ large, the sum in Eq.~(\ref{FN}) can be extended to infinity without significant error. However, we have to be careful about the periodicity in $q$ of the discretized version\footnote{This is similar to the difference between the delta function and its discretized version, the sinc function.}. That is, $F_q$ as given in Eq.~(\ref{FN}) is really a function of $q \mod 2$. Thus we first replace $q$ by $\bar{q}$, where $\bar{q}$ is the nearest to $q_0$ of the set of equivalent momenta, satisfying $|\bar{q}-q_0|\le 1$, to obtain the result
\bea\label{Fq}
F_q(x)= w\sqrt{\pi}\ e^{-\quarter\pi^2(\bar{q}-q_0)^2w^2/a^2+i\pi \bar{q} x/a} \vartheta_3\left( \pi\frac{x}{a}+\half i\pi^2 (\bar{q} - q_0)\frac{w^2}{a^2}\ ,\ e^{-\pi^2w^2/a^2}\right),
\eea
where $\vartheta_3(z,v)$ is the Jacobi elliptic theta function, which has the expansion\cite{AS}
\bea
\vartheta_3(z,v)=1+2\sum_{s=1}^\infty v^{s^2}\cos(2sz).
\eea
For a broad wave-packet, with $w\gg a$, the argument $v=e^{-\pi^2 w^2/a^2}$ is very small, so that $\vartheta_3(z,v)\approx 1$. Moreover the prefactor in Eq.~(\ref{Fq}) is also small except for the case $\bar{q}=q_0\equiv k_0a/\pi$.

We are now in a position to identify under what conditions the  Jordan associated functions are excited, that is, when $\beta_n$, as given by Eq.~(\ref{InIntegral}), is non-vanishing.
In order to obtain an appreciable value of $F_n$ the scaled offset momentum $q_0$ must be an integer $m$, with $n\equiv m$ mod 2. Now, however, there is the integral in the numerator to be considered. Given that the phase of $F_n(x)$ is approximately $e^{i\pi \bar{q} x/a}$ and that $I_n(y)$ has the expansion
\bea
I_n(y)=(\half y)^n \sum_{s=0}^\infty \frac{(\quarter y^2)^s}{s! \Gamma(n+s+1)},
\eea
where we recall that $y=y_0 e^{i\pi x/a}$, it is easily seen that the integral vanishes unless $m$ is negative and $n\le |m|$. In that case the Jordan functions $\vf_{|m|},\ \vf_{|m|-2},\ \dots$ are excited.  However, if $q_0$ is not a negative integer, no Jordan blocks are excited, and thus no linear growth is to be expected. For $q_0=0$ no Jordan function is excited because the ground-state level $n=0$ is non-degenerate. Thus no linear growth is expected in this case. Nor is it expected for the case $q_0=+1$. In the set-up chosen by Longhi in Ref.~\cite{op6}, on the other hand, the offset $q_0$ was taken to be -1, in which case $\vf_1$ is excited. Note the left-right asymmetry here: the Hamiltonian is not parity invariant, but only $PT$ invariant.

Figures 3 and 4 show the different propagation behaviour for a wide beam in the two cases $q_0=-1$ and $q_0=0$ respectively. The parameters are: $a=\pi$, $w=6\pi$ and  $V_0=2$. In the first case, where one Jordan mode $\vf_1$, is excited, the beam spreads out but does not split, as shown in Fig.~3(a). In Fig.~3(b) we show the different contributions to the maximum amplitude. The lower curve shows the contribution of the Jordan block sector only, which indeed rises linearly, as expected. However, the intermediate curve, the contribution of all other modes, mysteriously begins to decrease after an initial rise,  and the upper curve, which takes into account all the contributions,  exhibits the saturation first noted by Longhi\cite{op6}.

\begin{figure}[h!]
\begin{center}
\resizebox{6cm}{!}{\includegraphics{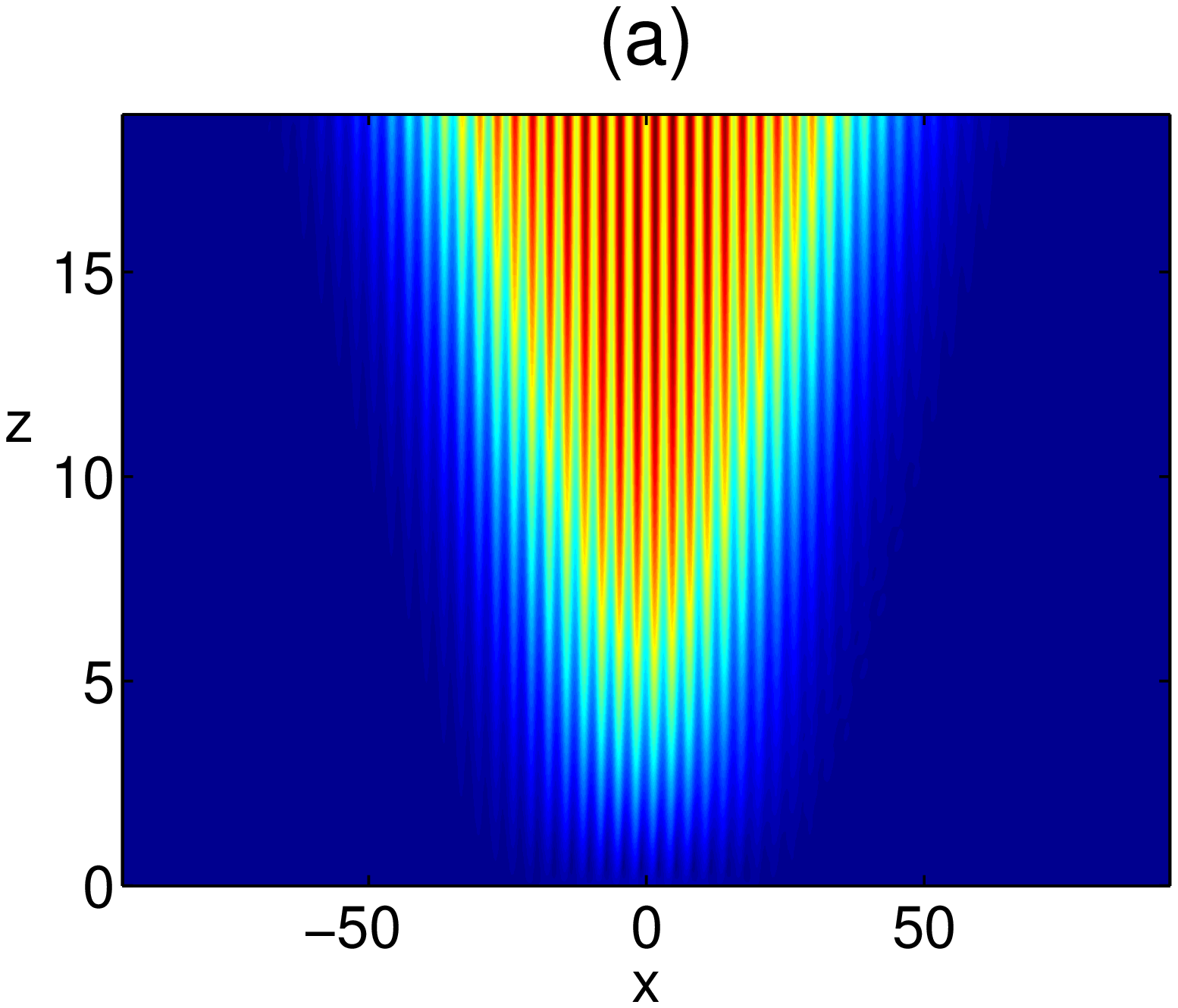}}\hspace{1cm} \resizebox{6cm}{!}{\includegraphics{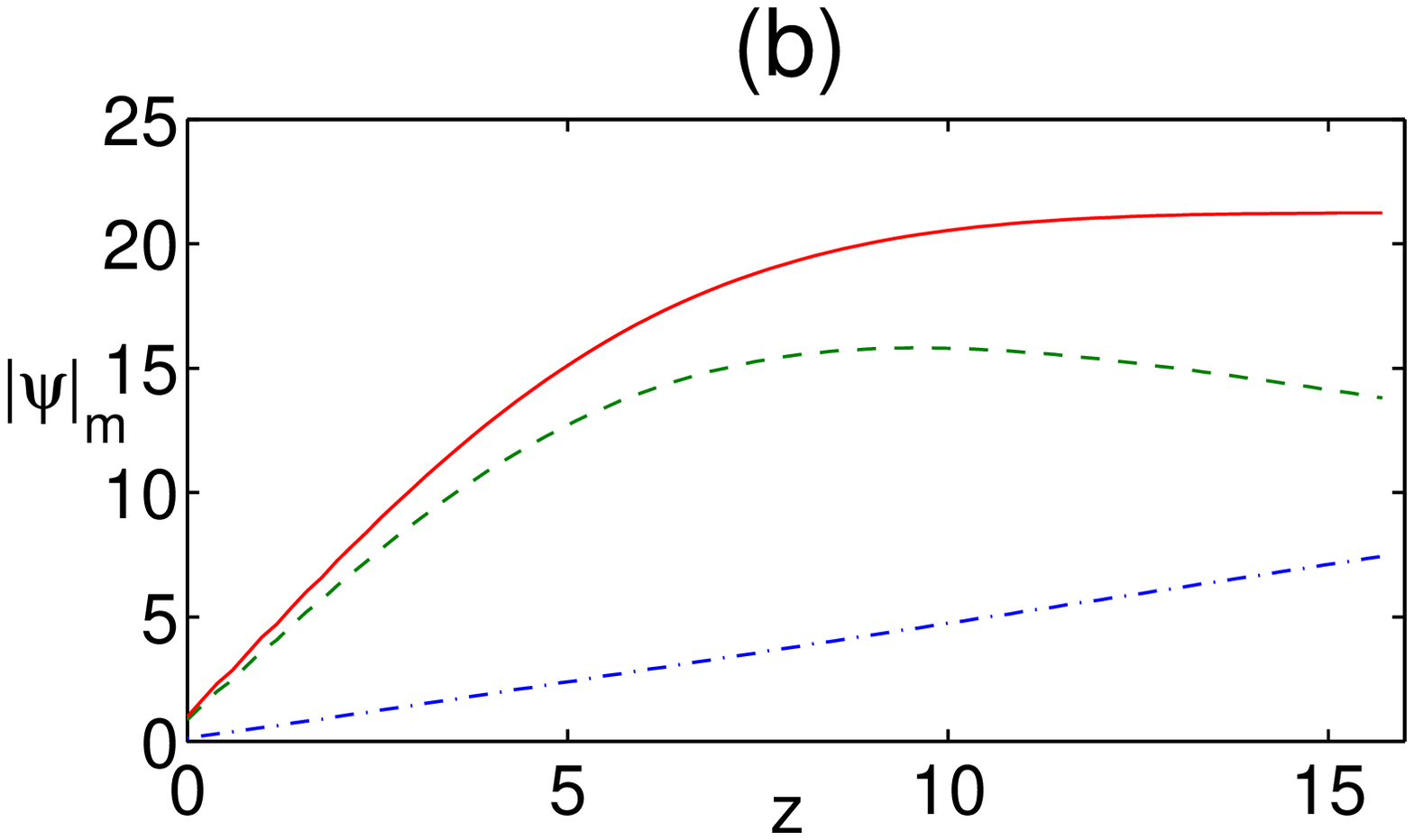}}
\caption{(color online) (a) $|\psi(x,z)|$ as a function of $x$, $z$. (b) Maximum value (red solid line) of $|\psi(x,z)|$ as a function of $z$. Blue dashed-dotted line: Jordan block contributions (second line of Eq.~(\ref{J+rest}) only). Green dashed line: other contributions only.
The parameters are: $a=\pi$, $V_0=2$, $w=6\pi$ and $q_0=-1$.}
\end{center}
\end{figure}

\begin{figure}[h!]
\begin{center}
{\includegraphics[width=6cm]{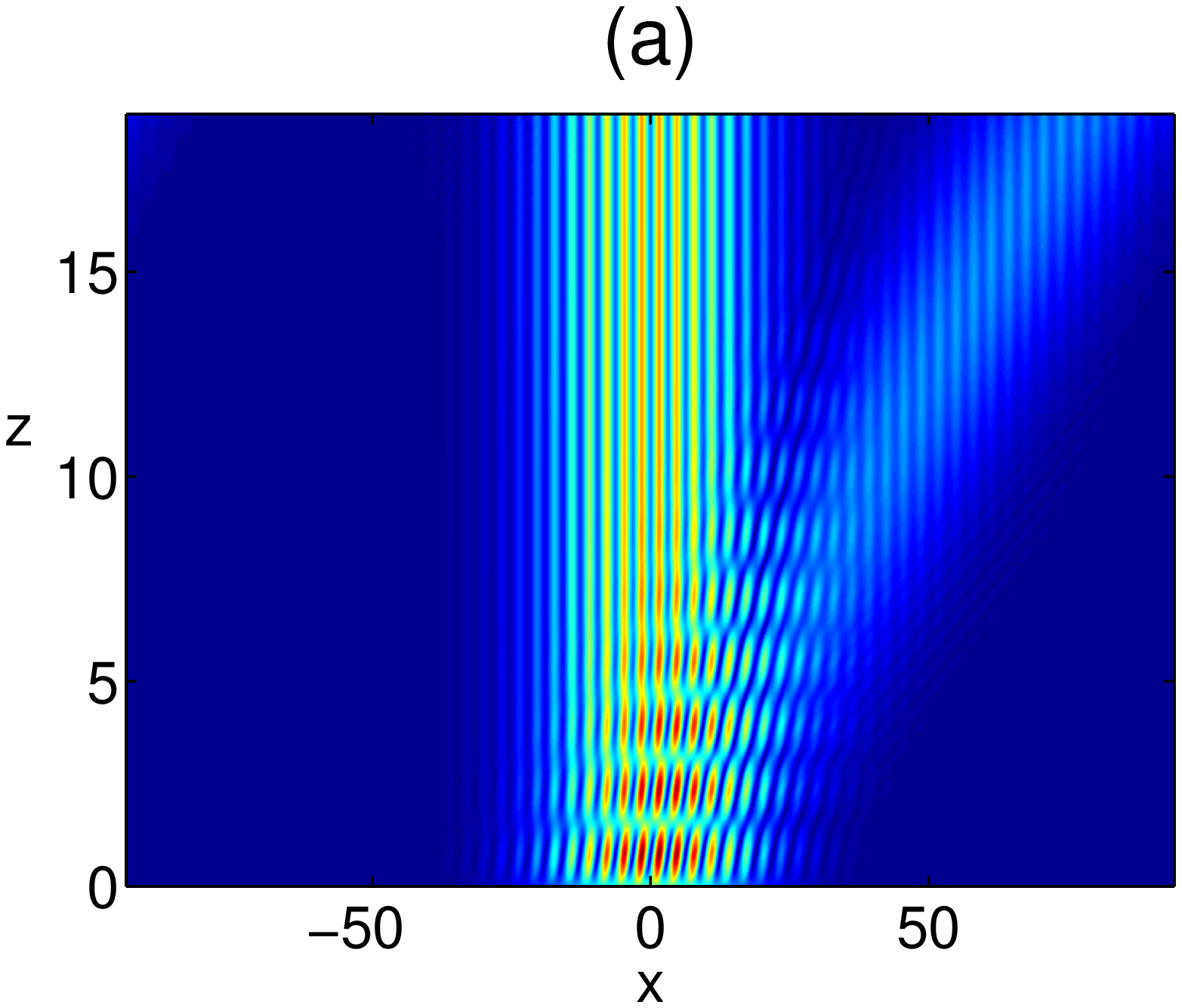}}\hspace{1cm}{\includegraphics[width=6cm]{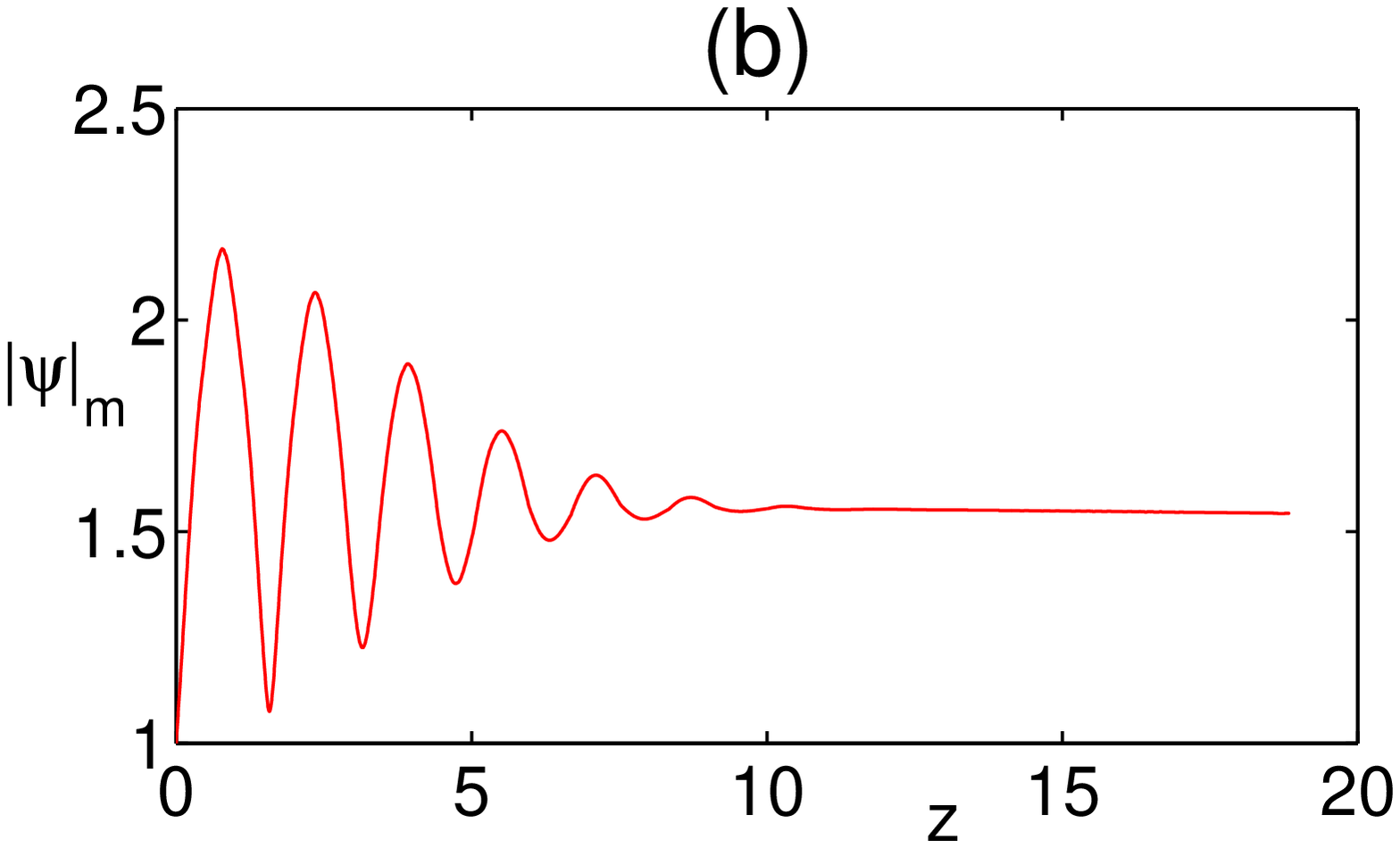}}
\caption{(color online) (a) $|\psi(x,z)|$ as a function of $x$, $z$. (b) Maximum value of $|\psi(x,z)|$ as a function of $z$.
The parameters are: $a=\pi$, $V_0=2$, $w=6\pi$ and $q_0=0$.}
\end{center}
\end{figure}

In the second case, where no Jordan mode is excited, the behaviour is quite different. The beam does not significantly spread, but instead splits into two, exhibiting the phenomenon of birefringence\cite{op3}, and the maximum amplitude, after some initial fluctuations, decreases slowly with $z$. The behaviour for the case when $q_0=1$ is similar.

\section{The mechanism of saturation}

In this section we investigate in detail how the contributions from the other (non-Jordan) eigenfunctions conspire to cancel out the explicit linear growth in $z$ coming from the Jordan sector. A priori such a cancellation seems highly unlikely, but since it does occur it must be because the two sets of contributions are in fact closely connected. Mathematically the cancellation cannot be complete, but can only happen only over a limited range in $z$. However, that is sufficient for the physical situation, because our lattice is finite in the $x$ direction, and at very large values of $z$ the beam will have encountered the edges of the lattice.

We now show that the contributions of the Jordan block and of the nearby eigenfunctions are indeed closely related. We will consider specifically the first case of the previous section, where the Jordan mode $\vf_1$ is excited.

Recall that
$\psi(x,0)\equiv f(x)$ is expanded as
\bea
\psi(x,0)=c_0 \psi_0(x)+\sum_{k\ne n\pi/a} c_k \psi_k(x)+\sum_{n >\ 0}(\alpha_n I_n(y)+\beta_n \chi_n(y)),
\eea
as in Eq.~(\ref{PTexpansion}).
The expression for $c_k$ is given in Eq.~(\ref{PTcoeffs}), in which the integration over $x$ can be reduced to the interval $[0,a]$
by exploiting the periodicity of the Bloch functions, to obtain
\bea\label{cknum}
c_k=\frac{\int_0^a dx \psi_{-k}(x) F_q(x)}{2N\int_0^a dx \psi_{-k}(x)\psi_k(x)},
\eea
where
\bea
F_q(x)\equiv \sum_{m=-N}^{N-1} e^{-i\pi m q}f(x+ma).
\eea
Similarly $\beta_n$ is given by
\bea
\beta_n = \frac{\int_0^a dx I_n(y) F_{n}(x)}{2N\int_0^a dx I_n(y) \chi_n(y)}.
\eea

The denominator of $c_k$ would vanish at $q=n$. In fact it turns out that it is precisely a sinc function:
\bea
\int_0^a dx \psi_{-k}(x)\psi_k(x)=a\ \mbox{sinc}(ka)=a\ \mbox{sinc}(q \pi).
\eea
The denominator of $\beta_n$ is proportional to the derivative of this previous denominator with respect to $k$. Thus
\bea
\frac{d}{dk}\int_0^a dx \psi_{-k}(x)\psi_k(x)&=&\frac{a}{\pi}\frac{d}{dq}\int_0^a dx I_{-q}(y)I_q(y)\cr
&=&\frac{a}{\pi}\int_0^a dx \left(I'_{-q}(y)I_q(y)+I_{-q}(y)I'_q(y)\right)\ .
\eea
But by differentiating the general relation $I_{-q}(y)=I_q(y)+(2/\pi)\sin{q\pi}\ K_q(y)$ and setting $q=1$ we find that
$I'_{-1}(y)=I'_1(y)-K_1(y)$. Thus, using the derivative definition of the Jordan associated function as\footnote{In fact this
differs from $\vf^{PI}$ as given in Eq.~(\ref{particular}) by a multiple of $I_q(y)$}
$\vf_k^{PI}=(1/(2k))d\psi_k/dk$, we see by reference to Eq.~(\ref{fixed}) that
\bea
\left.\frac{d}{dk}\int_0^a dx \psi_{-k}(x)\psi_k(x)\right|_{q=1}&=&4\frac{a}{\pi}
\int_0^a dx I_1(y)\left(\chi_1^{PI}(y)-\half K_1(y)\right)\cr
&=&4\frac{a}{\pi}\int_0^a dx I_1(y)\chi_1(y)\ .
\eea

The numerator of $\beta_n$ is a smooth continuation of the numerator of $c_k$, which from now on we denote as $\hat{c}_k$.
Moreover, $\hat{c}_k$ is highly peaked around $q=1$ (and also around $q=-1$), as is shown in Fig.~5.
In fact, near $q=1$ it is given by the Gaussian $\hat{c}_k\propto e^{-\pi^2\epsilon^2w^2/(4 a^2)}$,
where $\epsilon=q-1$, while the denominator is proportional to $\epsilon$.
\begin{figure}[h!]
\begin{center}
\resizebox{3in}{!}{\includegraphics{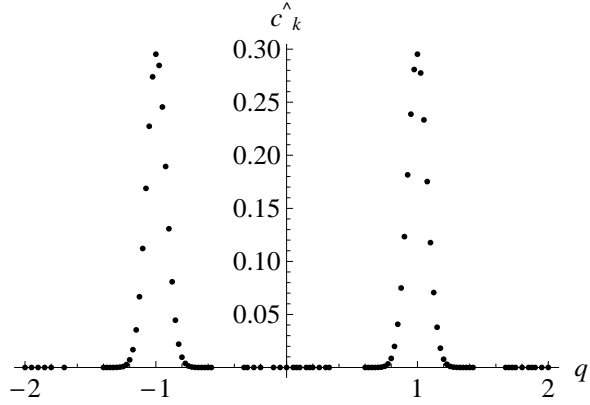}}
\caption{ The numerator $\hat{c}_k$ of the expansion coefficients $c_k$ in Eq.~(\ref{cknum}) for $N=40$, $a=\pi$, $w=6\pi$ and $V_0=2$.
At the points $q=\pm 1$, which are not included in the expansion, the figure gives instead the value of
the numerator of $\beta_1$}
\end{center}
\end{figure}

We are now in a position to examine the $z$-development of the Jordan contribution and the related contributions from nearby values of $q$. Recall that
$\psi(x,z)$ is given by
\beax
\psi(x,z)&=&c_0 \psi_0(x)+\sum_{k\ne n\pi/a} c_k \psi_k(x)e^{-ik^2z}\\
&&\hspace{1cm}+\sum_{n >\ 0}[(\alpha_n -iz\beta_n)I_n(y)+\beta_n \chi_n(y)]e^{-i(n^2\pi^2/a^2)x}.
\eeax

Thus the total contribution from the neighbourhood of $q=1$ and $q=-1$ is
\bea\label{flat}
\psi(x,z)_J\approx{\rm const} \times I_1(y)e^{-iz}\left[
z+\sum_{r=1}\left(\frac{a^2}{\pi^2\epsilon_r}\right)\sin\left(\frac{2\pi^2}{a^2}\epsilon_r z\right)e^{-\epsilon_r^2\pi^2(w^2/4+iz)/a^2}\right],
\eea
where $\epsilon_r=r/N$. In principle the upper limit of the sum is infinity, but in practice it can be taken less than $N$. Note that if $\epsilon$ were continuous the limit of the second term as $\epsilon\to 0$ would be twice the first term, which comes from the Jordan function. It turns out that although $|\psi(x,z)_J|$ does exhibit the expected linear behaviour in $z$ initially,  it subsequently$|\psi(x,z)_J|$ has an extremely wide and flat plateau before it eventually rises
again as $z$ approaches $N\pi$. This is illustrated in Fig.~6(a) for $N=40$. A hint of such a plateau-like behaviour can be seen in the simple function $z+\sin(2z)/2$, which is plotted in Fig.~6(b). However, in this case the plateau is much less pronounced.
\begin{figure}[h!]
\begin{center}
\resizebox{3in}{1.7in}{\includegraphics{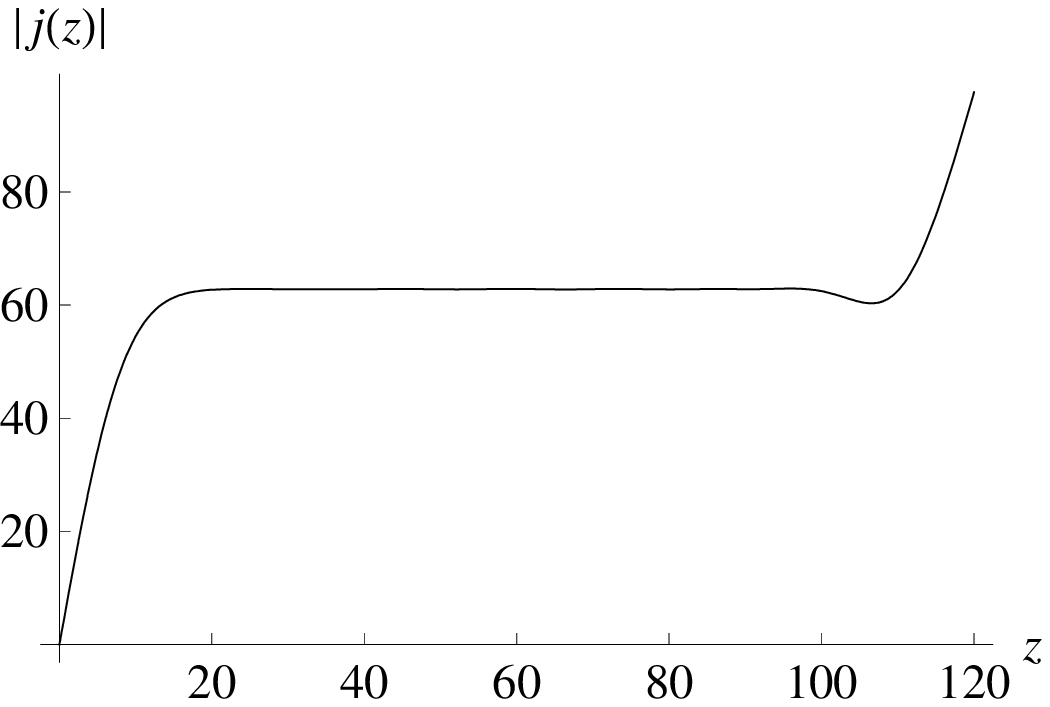}}\hspace{0.4in}\resizebox{3in}{1.7in}{\includegraphics{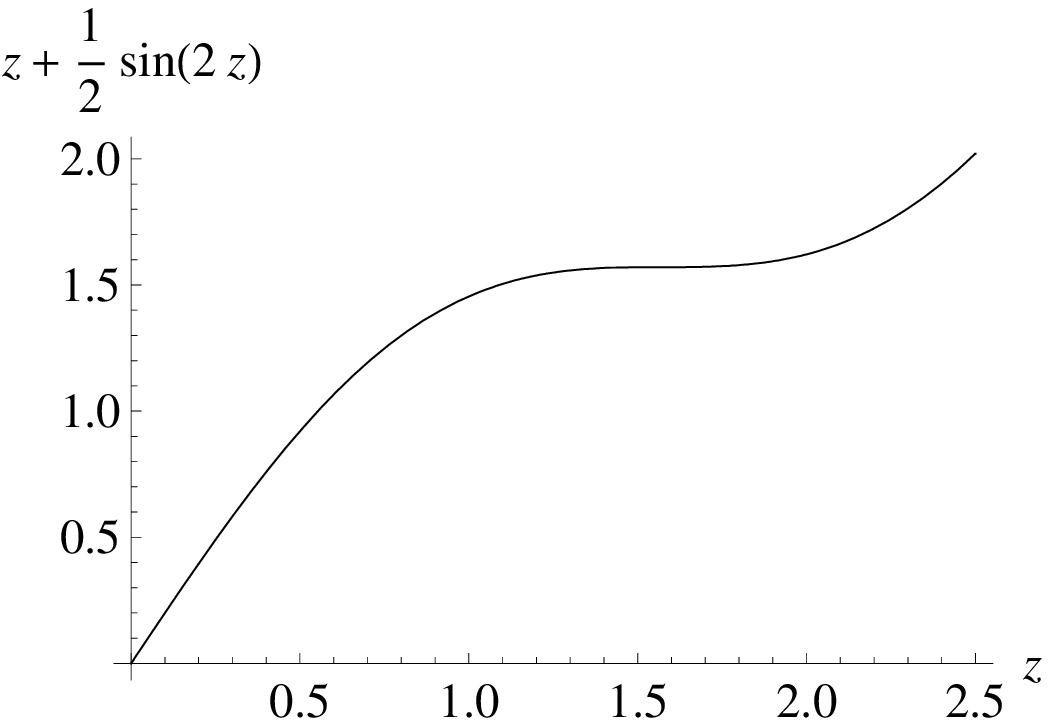}}
\hspace{1.2in}(a) \hspace{3in} (b)
\caption{(a) Modulus of the function in square brackets in Eq.~(\ref{flat}) as a function of $z$ for $N=40$, $a=\pi$ and $w=6\pi$. (b) The function $z+\sin(2z)/2$.}
\end{center}
\end{figure}

We can understand the extreme flatness of the plateau in Fig.~6(a) in terms of Jacobi $\vartheta_3$ functions.
For simplicity let us set $a=\pi$, so that we are analyzing the function
\bea
j(z)&\equiv&z+\sum_{r=1}^\infty\left(\frac{1}{\epsilon_r}\right)\sin(2\epsilon_r z)e^{-\epsilon_r^2/(w^2/4+iz)}\\
&\approx&z+\sum_{r=1}^\infty\left(\frac{1}{\epsilon_r}\right)\sin(2\epsilon_r z)e^{-4\epsilon_r^2/w^2}\nonumber
\eea
for the values of $z$ we are considering. While this is not itself a $\vartheta_3$ function, its derivative with respect to $z$ can be so expressed:
\bea
j'(z)&\approx& 1+2\sum_{r=1}^\infty\cos(2\epsilon_r z)e^{-\epsilon_r^2w^2/4}\cr
&=&1+2\sum_{r=1}^\infty\cos(2rz/N)e^{-r^2w^2/(4N^2)}\cr
&=&\vartheta_3\left(\frac{z}{N}, e^{-w^2/(4N^2)}\right)\ .
\eea
The behaviour of this $\vartheta$ function is not immediately apparent, since the second argument is of $O(1)$ for $w\ll 2N$. However, it can be made clear by using the alternative notation $\vartheta(z,q)=\vartheta_3(z|\tau)$, where $q=e^{i\pi\tau}$, and applying
Jacobi's imaginary transformation\cite{WW}, whereby
\bea
\vartheta_3(z|\tau)=(-i\tau)^{-\half}e^{-i\tau' z^2/(\pi\tau')}\vartheta_3(z\tau'|\tau'),
\eea
where $\tau'=-1/\tau$. This converts $j'(z)$ to
\bea\label{jp}
j'(z)=2\sqrt{\pi}\frac{N}{w} e^{-4z^2/w^2}\vartheta_3\left(\frac{4\pi i N}{w^2}z, e^{-4\pi^2N^2/w^2}\right).
\eea
In this form the second argument of $\vartheta_3$ is small, so that for moderate values of $z$ we can approximate $\vartheta_3$ by 1, in which case the behaviour is dominated by the preceding Gaussian, which rapidly falls from 1 to a very small value, corresponding to the plateau in $j(z)$. The Gaussian is eventually overwhelmed by the hyperbolic cosines occurring in the expansion of $\vartheta_3$, as must be the case, since $\vartheta_3$ is periodic in $z$. The plot of this function is given in Fig.~7.
\begin{figure}[h!]
\begin{center}
\resizebox{4in}{2in}{\includegraphics{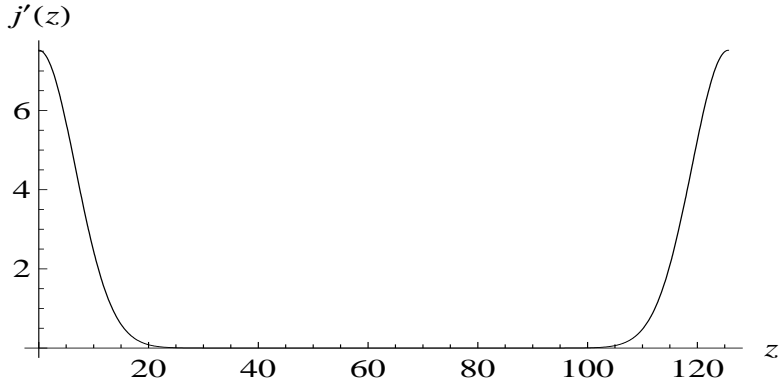}}
\caption{$j'(z)$ from Eq.~(\ref{jp}) versus $z$. The parameters are the same as in Fig.~6.}
\end{center}
\end{figure}

\section{Conclusions}
On general grounds one expects linear growth of the amplitude $\psi(x,z)$ at the symmetry-breaking threshold $\lambda=1$, due to the degeneracy of a subset of the eigenfunctions at this point and the consequent development of Jordan blocks. However, it has been observed numerically\cite{op6} that this growth becomes saturated at large $z$, at least for wide input beams. We have been able to explain this saturation phenomenon by analyzing in detail the separate contributions from the Jordan blocks and the contributions from the nearby Bloch functions in which they are embedded.

For the particular potential considered here, the Bloch eigenfunctions are associated Bessel functions of the first kind, $I_q(y)$, and we were able to explicitly construct the associated Jordan functions at the exceptional points $q=n$. Hence we were in a position to use the analogue of the method of stationary states to generate the $z$ dependence. In this method we were able to isolate the separate contributions of the Jordan blocks from the other non-degenerate states with normal $z$ dependence generated by multiplying each eigenfunction by its appropriate $z$-dependent phase. We found that, in cases where the associated Jordan functions are excited (Fig.~3) the explicit linear increase of the Jordan-block contributions is precisely compensated by a linear decrease of the contribution of the non-degenerate states, which of course have no {\sl explicit} linear $z$-dependence. In cases where the Jordan associated functions are not excited (Fig.~4), there is no initial linear increase, but rather a rapid oscillatory behaviour followed by a very slow decrease. The topology of the beams in these two cases is markedly different. In the first case, although the maximum amplitude becomes constant, the beam spreads laterally in a linear fashion, and hence the total power grows linearly, with the beam taking energy from the lattice. In the second case the total power turns out to be constant for large $z$, with the slow decrease in the maximum amplitude being matched by the slow broadening of the individual beams.

We then examined in detail the mechanism of saturation, which is only possible because of the close relation between the contribution of the Jordan block and the contributions of the Bloch functions in which it is embedded. With the aid of a certain amount of approximation we were able to write the respective contributions in the relatively simple form of Eq.~(\ref{flat}) and to express its $z$-derivative as a $\vartheta_3$ function. This $\vartheta_3$ function encodes the initial increase and the subsequent extremely wide and flat plateau. As a mathematical function it also encodes further steps and plateaux for larger $z$, but these are not physically relevant because they correspond to values of $x$ outside the range of the finite lattice.

\acknowledgments{We are grateful to Prof. R.~J.~Rivers for extremely useful comments. EMG is supported by an Imperial College Junior Research Fellowship. }

\section*{APPENDIX}
Since the references for Jordan associated functions are not readily accessible, we give here a brief outline of their main properties. They are the analogue of the Jordan associated vectors that occur in simple matrix eigenvalue problems of the form
\bea
(H-\lambda)u=0
\eea
where $H$ is non-Hermitian.  For Hermitian problems, while two eigenvalues $\lambda_1$ and $\lambda_2$ may become degenerate at a particular critical value of a parameter in the Hamiltonian, there remain two distinct eigenvectors. However, in the non-Hermitian case it is possible that the eigenvectors also become degenerate: $u_1=u_2$.  The reduction in the number of eigenvectors at such a point means that the set of eigenvectors no longer forms a complete basis. However, the basis can be completed by inclusion of the Jordan associated vector $v_1$ (generalized eigenfunction), defined by the generalized eigenvalue equation
\bea
(H-\lambda_1)v_1=u_1\ .
\eea
Of course this associated vector is not uniquely defined: to it may be added any multiple of $u_1$.

The archetypical example is given by the non-Hermitian Jordan matrix
\bea\label{M}
M=\left(\begin{array}{cc}\lambda & 1\\ 0 & \lambda\end{array}\right),
\eea
 which has
the single eigenvalue $\lambda$ and only one eigenvector $u=(1,0)$. The independent vector $v\equiv(0,1)$ needed to complete the basis
is indeed a solution (undetermined up to a multiple of $u$) of the generalized eigenvalue equation
\bea\label{v}
(M-\lambda)v=u\ .
\eea
The eigenvectors of a non-Hermitian matrix are not orthogonal in the usual sense. Instead one needs to introduce the left eigenvectors $u_L$, satisfying $\tilde{u}_L(H-\lambda) =0$, which are different from the usual (right) column vectors satisfying $(H-\lambda)u_R=0$. The orthogonality is then between left and right eigenvectors: $(\tilde{u}_L)_1 (u_R)_2=0$ for distinct eigenvalues $\lambda_1$ and $\lambda_2$. When the eigenvectors become degenerate, they are self-orthogonal. This is exemplified by the Jordan matrix $M$ in Eq.~(\ref{M}), whose left and right eigenvectors are $u_L=(0,1)$ and $u_R\equiv u=(1,0)$, respectively, which are indeed orthogonal.

It is easy to show in general that the associated vectors $v_n$ are orthogonal in the same sense to eigenvectors $u_m$ and associated vectors $v_m$ belonging to different eigenvalues. In the event that $H$ has a $PT$ symmetry, for some definition of the reflection operator $P$, all of these overlaps can instead be expressed in terms of right eigenvectors alone with the aid of the $PT$ metric.

The continuum analogue of this situation is the eigenvalue problem
\bea
(H-E_n)\psi_n=0
\eea
where $H$ is a non-Hermitian differential operator with eigenvalue $E_n$ and (right) eigenfunction $\psi_n$.
Again, it is possible that, at a particular critical value of a parameter in $H$ two
eigenvalues $E_m$ and $E_n$, and also the corresponding eigenfunctions $\psi_m$ and $\psi_n$,  become degenerate.
At that point the single eigenfunction becomes self-orthogonal with respect to the metric defined by $H$, which, for a $PT$-symmetric problem, is the $PT$ metric.

 The reduction in the number of eigenfunctions at such a point again means that the set of eigenfunctions no longer forms a complete basis. However, in complete analogy with Eq.~(\ref{v}), the basis can be completed by inclusion of the Jordan associated functions $\vf_n$ (generalized eigenfunctions), defined by the generalized eigenvalue equation
\bea
(H-E_n)\vf_n=\psi_n\ .
\eea
This is precisely how the Jordan functions were introduced in Eq.~(\ref{Jordan}). They are defined only up to multiples of solutions of the homogeneous equation and are guaranteed to be orthogonal, using the $PT$ metric, to each other and to eigenfunctions belonging to other eigenvalues. We found it necessary to exploit this freedom in Eq.~(\ref{fixed}) in order to ensure that the $\vf_n$ satisfy the appropriate boundary conditions.

An alternative definition of the Jordan functions is as derivatives of the eigenfunctions with respect to the energy\cite{Mondragon}. Thus, differentiating
the eigenvalue equation $(H-E)\psi=0$ with respect to $E$ we get precisely
\bea
(H-E)\frac{d\psi}{dE}-\psi=0\ ,
\eea
so that we may identify $\vf$ with $d\psi/dE$, again modulo solutions of the homogeneous equation.

In our present problem this latter definition leads to extremely simple formulae for the functions $\chi_n$. For example, using Eq.~(9.6.44) of Ref.~\cite{AS} it yields  $\chi_1(y)=-I_0(y)/(2y)$, which is easily seen to be a solution of Eq.~(\ref{Jordan}) with the correct periodicity.


\begin{thebibliography}{99}
\bibitem{BB} C.~M.~Bender and S.~Boettcher, Phys.~Rev.~Lett. {\bf 80},
5243 (1998).
\bibitem{CMBR} C.~M.~Bender, Contemp.~Phys. {\bf 46}, 277 (2005);
Rep.~Prog.~Phys. {\bf 70}, 947 (2007).
\bibitem{AMR} A.~Mostafazadeh, Int. J. Geom. Meth. Mod. Phys. {\bf 7}, 1191 (2010).
\bibitem{BBJC} C.~M.~Bender, D.~C.~Brody and H.~F.~Jones,
Phys.~Rev.~Lett. {\bf 89}, 270401 (2002) ; 92, 119902(E) (2004).
\bibitem{BBJQ} C.~M.~Bender, D.~C.~Brody and H.~F.~Jones,
Phys.~Rev.~D {\bf 70}, 025001 (2004) ; 71, 049901(E) (2005).
\bibitem{AMh} A.~Mostafazadeh, J.~Math.~Phys. {\bf 43}, 205 (2002);
J.~Phys.~A {\bf 36}, 7081 (2003).
\bibitem{op1} R.~El-Ganainy, K.~G.~Makris, D.~N.~Christodoulides and Z.~H.~Musslimani, Optics Letters {\bf 32}, 2632 (2007).
\bibitem{op2} Z.~Musslimani, K.~G.~Makris, R.~El-Ganainy and D.~N.~Christodoulides,  Phys.~Rev.~Lett. {\bf 100}, 030402 (2008).
\bibitem{op3} K.~Makris, R.~El-Ganainy, D.~N.~Christodoulides and Z.~Musslimani, Phys.~Rev.~Lett. {\bf 100}, 103904 (2008).
\bibitem{op4} S.~Klaiman, U.~G\"unther and N.~Moiseyev, Phys.~Rev.~Lett. {\bf 101} 080402 (2008).
\bibitem{op5} A.~Guo, G.~J.~Salamo, D.~Duchesne, R.~Morandotti, M.~Volatier-Ravat, V.~Aimez, G.~A.~Siviloglou and D.~N.~Christodoulides, Phys.~Rev.~Lett. {\bf 103}, 093902 (2009).
\bibitem{op6} S.~Longhi, Phys.~Rev.~A {\bf 81}, 022102 (2010).
\bibitem{op7} K.~Makris, R.~El-Ganainy, D.~N.~Christodoulides and Z.~Musslimani, Phys.~Rev.~A {\bf 81}, 063807 (2010).
\bibitem{op8} C.~R\"uter, K~G.~Makris, R.~El-Ganainy, D.~N.~Christodoulides, M.~Segev and D.~Kip, Nature Physics {\bf 6}, 192 (2010).
\bibitem{op9} H.~Ramezani, T.~Kottos, R.~El-Ganainy and D.~Christodoulides, Phys.~Rev.~A {\bf 82}, 043803 (2010).
\bibitem{Berry98} M.~V.~Berry, J. Phys. A  {\bf 31}, 3493 (1998).
\bibitem{Kottos_uni} M.~C.~Zengh, D.~N.~Christodoulides, R.~Fleischmann and T.~Kottos, Phys.~Rev.~A  {\bf 82}, 010103(R)  (2010).
\bibitem{Longhi_bloch} S.~Longhi, Phys.~Rev.~Lett. {\bf 103}, 123601 (2009).
\bibitem{Longhi_dl} S.~Longhi, Phys.~Rev.~B {\bf 80}, 235102 (2009).
\bibitem{Kulishov} M.~Kulishov, J.~M.~Laniel, N.~B\'elanger, J.~Aza\~na and D.~Plant, Optics Express, {\bf 13}, 3068 (2005).
\bibitem{nhbh} E.~M.~Graefe, H.~J.~Korsch and A.~E.~Niederle,  Phys.~Rev.~Lett. {\bf 101}, 150408 (2008); E.~M.~Graefe, H.~J.~Korsch and A.~E.~Niederle,
  Phys.~Rev.~A  {\bf 82} 013629 (2010).
\bibitem{Midya} B.~Midya, B.~Roy and R.~Roychoudhury, Phys.~Lett. A {\bf 374}, 2605 (2010).
\bibitem{Heiss} W.~D.~Heiss, European Physical Journal D {\bf 60}, 257 (2010).
\bibitem{Kato} T.~Kato, {\it Perturbation theory for linear operators}
(Springer, Berlin,1966).
\bibitem{Keldysh} M.~V.~Keldysh, Russ.~Math.~Surv. {\bf 26}, 15, 1971.
\bibitem{AS} M.~Abramowitz and I.~A.~Stegun,  {\sl Handbook of Mathematical Tables}, Dover, New York 1970.
\bibitem{WW} E.~T.~Whittaker and G.~N.~Watson,  A Course in Modern Analysis, 4th ed., Cambridge University Press 1990.
\bibitem{Mondragon} E.~Hern\'andez, A.~J\'auregui and A.~Mondrag\'on, Phys.~Rev.~A {\bf 67}, 022721 (2003).
\end{thebibliography}
\end{document}